\documentclass[12pt]{article}
\usepackage{url}
\usepackage{graphicx}
\usepackage{float}
\usepackage{sansmath}
\usepackage{xcolor}
\usepackage{setspace}
\usepackage{amsmath}
\usepackage{natbib}
\usepackage{authblk}
\usepackage[section]{placeins}

\begin{document}
\title{Forecasting football matches by predicting match statistics}
\author{Edward Wheatcroft}
\affil{\footnotesize London School of Economics and Political Science, Houghton Street, London, United Kingdom, WC2A 2AE.}
\maketitle
\noindent

\begin{abstract}
This paper considers the use of observed and predicted match statistics as inputs to forecasts of the outcomes of football matches. It is shown that, were it possible to know the match statistics in advance, highly informative forecasts of the match outcome could be made. Whilst, in practice, match statistics are never available prior to the match, this leads to a simple philosophy.  If match statistics can be predicted pre-match, and if those predictions are accurate enough, it follows that informative match forecasts can be made. It is shown that Generalised Attacking Performance (GAP) ratings, introduced in a recent paper, provide a suitable methodology for predicting match statistics in advance and that they are informative enough to provide information beyond that reflected in the odds. A long term and robust gambling profit is demonstrated when the forecasts are combined with two betting strategies.
\end{abstract}

\section{Introduction}
Quantitative analysis of sports is a rapidly growing discipline with participants, coaches, owners, as well as gamblers, increasingly recognising its potential in gaining an edge over their opponents.  This has naturally led to a demand for information that might allow better decisions to be made.  Association football (hereafter football) is the most popular sport globally and, although, historically, the use of quantitative analysis has lagged behind that of US sports, this is slowly changing. Gambling on football matches has also grown significantly in popularity in recent decades and this has contributed to increased demand for informative quantitative analysis. \par

Today, in the most popular football leagues globally, a great deal of match data are collected.  Data on the location and outcome of every match event can be purchased, whilst free data are available including match statistics such as the numbers of shots, corners and fouls by each team.  This creates huge potential for those able to process the data in an informative way. This paper focuses on probabilistic prediction of the outcomes of football matches, i.e. whether the match ends with a home win, a draw or an away win.  A probabilistic forecast of such an event simply consists of estimated probabilities placed on each of the three possible outcomes.  Statistical models can be used to incorporate information into probabilistic forecasts. \par

The basic philosophy of this paper is as follows.  Suppose, somehow, that certain match statistics, such as the number of shots or corners achieved by each team, were available in advance of kickoff.  In such a case, it would be reasonable to expect to be able to use this information to create informative forecasts and it is shown that this is the case.  Obviously, in reality, this information would never be available in advance.  However, if one can use statistics from past matches to \emph{predict} the match statistics before the match begins, and those predictions are accurate enough, they can be used to create informative forecasts of the match outcome.  The quality of the forecast is then dependent both on the importance of the match statistic itself and the accuracy of the pre-match prediction of that statistic. \par 

In this paper, observed and predicted match statistics are used as inputs to a simple statistical model to construct probabilistic forecasts of match outcomes.  First, observed match statistics in the form of the number of shots on target, shots off target and corners, are used to build forecasts and are shown to be informative.  The observed match statistics are then replaced with predicted statistics calculated using Generalised Attacking Performance (GAP) Ratings, a system introduced in a recent paper which uses past data to estimate the number of defined measures of attacking performance a team can be expected to achieve in a given match (\cite{Wheatcroft2020}).  Whilst, unsurprisingly, it is found that predicted match statistics are less informative than observed statistics, it is found that they can still provide useful information for the construction of the forecasts. It is shown that a robust profit can be made by constructing forecasts based on predicted match statistics and using them alongside two different betting strategies. \par

For much of the history of sports prediction, ratings systems in a similar vein to the GAP rating system used in this paper, have played a key role. Probably the most well known is the Elo rating system which was originally designed to produce rankings for chess players but has a long history in other sports (\cite{elo1978rating}).  The Elo system assigns a rating to each player or team which, in combination with the rating of the opposition, is used to estimate the probability of each possible outcome.  The ratings are updated after each game in which a player or team is involved. A weakness of the original Elo rating system is that it does not estimate the probability of a draw.  As such, in sports such as football, in which draws are common, some additional methodology is required to estimate that probability. \par

Elo ratings are in widespread use in football and have been demonstrated to perform favourably with respect to other rating systems (\cite{hvattum2010using}). Since 2018, Fifa has used an Elo rating system to produce its international football world rankings (\cite{Fifa_elo}).  Elo ratings have also been applied to a wide range of other sports including, among others, Rugby League (\cite{carbone2016rugby}) and video games (\cite{suznjevic2015application}). The website \url{fivethirtyeight.com} produces probabilities for NFL (\cite{538}) and NBA (\cite{538NBA}) based on Elo ratings. Another limitation of the Elo rating system is that it does not account for the \emph{size} of a win.  This means that a team's ranking after a match would be the same after either a narrow or convincing victory. Some authors have adapted the system to account for the margin of victory (see, for example, \cite{lasek2013predictive} and \cite{sullivan2016improving}). \par

The original Elo rating system assigns a single rating to each participating team or player, reflecting its overall ability. This does not directly allow for a distinction between the performance of a team in its home or away matches. Typically, some adjustment to the estimated probabilities is made to account for home advantage.  Other ratings systems distinguish between home and away performances.  One system that does this is the pi-rating system in which a separate home and away rating is assigned to each team (\cite{constantinou2013determining}).  The pi-rating system also takes into account the winning margin of each team, but this is tapered such that the impact of additional goals on top of already large winning margins is lower than that of goals in close matches. \par
 
The GAP rating system, introduced in \cite{Wheatcroft2020} and used throughout this paper, differs from both the Elo rating and the pi-rating systems in that, rather than producing a single rating, each team is assigned a separate attacking and defensive rating both for its home and away matches.  This results in a total of 4 ratings per team. The approach of assigning attacking and defensive ratings has been taken by a large number of authors. An early example is \cite{maher1982modelling} who assigned fixed ratings to each team and combined them with a Poisson model to estimate the number of goals scored.  They did not use their ratings to estimate match probabilities but \cite{dixon1997modelling} did so using a similar approach.  Combined with a value betting strategy, they were able to demonstrate a significant profit for matches with a large discrepancy between the estimated probabilities and the probabilities implied by the odds. \cite{dixon2004value} modified the Dixon and Coles model and were able to demonstrate a profit using a wider range of published bookmaker odds. \cite{doi:10.1111/1467-9884.00243} defined a Bayesian model for attacking and defensive ratings, allowing them to vary over time. Other examples of systems that use attacking and defensive ratings include \cite{karlis2003analysis}, \cite{lee1997modeling} and \cite{baker2015time}. \par

The use of ratings systems naturally leads to the question of how to translate them into probabilistic forecasts.  One of two approaches is generally taken.  The first is to model the number of goals scored by each team using Poisson or Negative Binomial regression with the ratings of each team used as predictor variables.  The match probabilities are then estimated through simulation. The second approach is to  predict the probability of each match outcome directly using methods such as logistic regression. There is little evidence to suggest a major difference in the performance of the two approaches (\cite{goddard2005regression}).  In this paper, the latter approach is taken, specifically in the form of ordinal logistic regression. \par

The idea that match statistics might be more informative than goals in terms of making match predictions has become more widespread in recent years.  The rationale behind this view is that, since it is difficult to score a goal and luck often plays an important role, the number of goals scored by each team might be a poor indicator of the events of the match.  It was shown by \cite{Wheatcroft2020} that, in the over/under 2.5 goals market, the number of shots and corners provide a better basis for probabilistic forecasting than goals themselves.  Related to this is the concept of `expected goals' which is playing a more and more important role in football analysis. The idea is that the quality of a shot can be measured in terms of its likelihood of success. The expected goals from a particular shot corresponds to the number of goals one would `expect' to score by taking that shot. The number of expected goals by each team in a match then gives an indication of how the match played out in terms of efforts at goal.  Several academic papers have focused on the construction of expected goals models that take into account the location and nature of a shot (\cite{eggels2016expected,rathke2017examination}). \par

\section{Background} \label{section:background}
\subsection{Betting odds}
In this paper, betting odds are used both as potential inputs to models and as a tool with which to demonstrate profit making opportunities.  Decimal, or `European Style', betting odds are considered throughout.  Decimal odds simply represent the number by which the gambler's stake is multiplied in the event of success. For example, if the decimal odds are $2$, a \pounds 10 bet on said event would result in a return of $2 \times \pounds 10 = \pounds 20$. \par

Another useful concept is that of the `odds implied' probability.  Let the odds for the $ith$ outcome of an event be $O_{i}$. The odds implied probability is simply defined as the multiplicative inverse, i.e. $r_{i}=\frac{1}{O_{i}}$. For example, if the odds on two possible outcomes of an event (e.g. home or away win) are $O_{1}=3$ and $O_{2}=1.4$, the odds implied probabilities are $r_{1}=\frac{1}{3} \approx 0.33$ and $R_{2}=\frac{1}{1.4} \approx 0.71$.  Note how, in this case, $r_{1}$ and $r_{2}$ add to more than one.  This is because, whilst conventionally, probabilities over a set of exhaustive events should add to one, this need \emph{not} be the case for odds implied probabilities.  In fact, usually,  the sum of odds implied probabilities for an event will exceed one. The excess represents the bookmaker's profit margin or the `overround' which is formally defined as
\begin{equation}
\pi=\left(\sum_{i=1}^{m} \frac{1}{O_{i}}\right)-1.
\end{equation}
Generally, the larger the overround, the more difficult it is for a gambler to make a profit since the return from a winning bet is reduced. \par

\subsection{Data} \label{section:data}
This paper makes use of the large repository of data available at \url{www.football-data.co.uk}, which supplies free match-by-match data for 22 European Leagues. For each match, statistics are given including, among others, the number of shots, shots on target, corners, fouls and yellow cards. Odds data from multiple bookmakers are also given for the match outcome market, the over/under 2.5 goal market and the Asian Handicap match outcome market.  For some leagues, match statistics are available from the 2000/2001 season onwards. For others, these are available for later seasons.  Therefore, since the focus of this paper is forecasting using match statistics, only matches from the 2000/2001 season onwards are considered.  The data used in this paper are summarised in table~\ref{table:Leagues_available} in which, for each league, the total number of matches since 2000/2001, the number of matches in which shots and corner data are available and the number treated as eligible for betting is shown. The meaning of `eligible' is explained in more detail in section~\ref{section:gap_ratings} but relates to matches both in which match statistics are available and the home team has played at least six matches and has at least six matches left to play in the season.  All leagues include data up to and including the end of the 2018/19 season. \par

\begin{table}[!htb]
\begin{center}
\begin{tabular}{|l|r|r|r|}
\hline
League & No. matches & Match data available & Eligible \\ 
\hline
Belgian Jupiler League & 5090 & 480 & 288 \\
English Premier League & 7220 & 7220 & 4678 \\ 
English Championship & 10488 & 10487 & 7341 \\
English League One & 10463 & 10461 & 7315 \\  
English League Two & 10463 & 10459 & 6921 \\
English National League & 7040 & 5384 & 3951 \\  
French Ligue 1 & 7040 & 4907 & 3350 \\
French Ligue 2 & 7220 & 760 & 519 \\
German Bundesliga & 5786 & 5480 & 1837 \\ 
German 2.Bundesliga & 5670 & 1059 & 304 \\
Greek Super League & 4634 & 477 & 287 \\
Italian Serie A & 6894 & 5275 & 3599 \\
Italian Serie B & 8502 & 803 & 559 \\
Netherlands Eredivisie & 5814 & 612 & 396 \\
Portugese Primeira Liga & 5286 & 612 & 396 \\
Scottish Premier League & 4308 & 4306 & 2777 \\
Scottish Championship & 3334 & 530 & 238 \\
Scottish League One & 3335 & 540 & 239 \\
Scottish League Two & 3328 & 539 & 238 \\
Spanish Primera Liga & 7190 & 5290 & 3610 \\
Spanish Segunda Division & 8778 & 903 & 645 \\
Turkish Super lig & 5779 & 612 & 396 \\
\hline
Total & 143672 & 77196 & 49884 \\
\hline
\end{tabular}
\caption{Data used in this paper.}
\label{table:Leagues_available}
\end{center}
\end{table}

\section{Methodology}

\subsection{GAP ratings}
The Generalised Attacking Performance (GAP) rating system, introduced by \cite{Wheatcroft2020}, is a rating system for assessing the attacking and defensive strength of a sports team with relation to a particular measure of attacking performance such as the number of shots or corners.  For a particular given measure of attacking performance, each team in a league is given an attacking and a defensive rating, both for home and away matches.  An attacking GAP rating can be interpreted as an estimate of the number of defined attacking plays the team can be expected to achieve against an average team in the league, whilst its defensive rating can be interpreted as an estimate of the number of attacking plays it can be expected to concede against an average team. The GAP ratings of the ith team in a league are denoted as follows:

\begin{itemize}
\item $H_{i}^{a}$ - Home attacking GAP rating of the ith team in a league.
\item $H_{i}^{d}$ - Home defensive GAP rating of the ith team in a league.
\item $A_{i}^{a}$ - Away attacking GAP rating of the ith team in a league.
\item $A_{i}^{d}$ - Away defensive GAP rating of the ith team in a league.
\end{itemize}
\noindent A team's ratings are updated after each match it plays. Details of how GAP ratings evolve over time are given in the appendix.  \par

In \cite{Wheatcroft2020}, the aim was to build probabilistic forecasts of whether the number of goals in a match would exceed $2.5$.  To do this, the sum of the attacking and defensive GAP ratings of the two teams was used as a predictor variable in a logistic regression model.  In this paper, in which forecasts of the match outcome are sought, a different approach is needed. In a match with the ith team at home to the jth team, the estimated total number of defined attacking plays by the home team can be estimated with $\hat{S}_{h}=\frac{H_{i}^{a}+A_{j}^{d}}{2}$ whilst the number of attacking plays by the away team can be estimated with $\hat{S}_{a}=\frac{A_{j}^{a}+H_{i}^{d}}{2}$, that is the estimated number of defined attacking plays by the home team is the average of its home attacking rating and the away team's away defensive rating and the estimated number of defined attacking plays by the away team is the average of its away attacking rating and the home team's home defensive rating.  The predicted difference in the number of defined attacking plays made by the two teams is therefore given by $\hat{S}_{h}-\hat{S}_{a}$ and it is this quantity that is of interest in the prediction model. \par

\subsection{Parameter Selection for GAP ratings} \label{section:parameter_selection}
GAP ratings are determined by three parameters: one which determines the `learning rate', that is the importance of a team's previous match in its ratings, one that determines the effect of home matches on a team's away ratings and one that determines the effect of an away match on a team's home ratings. In \cite{Wheatcroft2020}, parameter selection was performed with the aim of optimising the skill of the resulting probabilistic forecasts.  In this paper, a different approach is taken.  Here, the parameters are selected using least-squares minimisation, with the aim of minimising the mean squared error between the estimated number of attacking plays and the observed number. That is, the function to be minimised is
\begin{equation}
f=\sum_{i=1}^{N} (S_{h}-\hat{S_{h}})^{2}+(S_{a}-\hat{S_{a}})^{2}
\end{equation}
where $S_{h}$ and $S_{a}$ are the observed numbers of attacking plays for the home and away team respectively and $\hat{S_{h}}$ and $\hat{S_{a}}$ are the predicted numbers.  This approach is taken with the aim that, as much as possible, the GAP ratings resemble estimates of the number of attacking plays made by each team.  Additionally, this allows combinations of GAP ratings produced under different measures of attacking performance to be applied simultaneously (in \cite{Wheatcroft2020}, only one set of GAP ratings were used at a time in the construction of the forecasts). \par

\subsection{Constructing Probabilistic Forecasts}
The nature of football matches is that the three possible outcomes can be considered to be `ordered'. Clearly, a home win is `closer' to a draw than it is to an away win.  As such, an appropriate model for predicting the probability of each outcome is ordinal logistic regression and this is the approach taken here. \par

Ordinal logistic regression is defined as follows.  Let $O_{1},...,O_{j}$ be $j$ ordered outcomes with probabilities $p_{1},...,p_{j}$ where $\sum_{i}^{j} p_{i} = 1$.  The model is parameterised as
\begin{equation} \label{eq:ord}
\log(\frac{p(Y \geq j)}{p(Y<j)})=\hat{\alpha}+\sum_{i=1}^{K} \hat{\beta_{i}}V_{i}
\end{equation}
where $V_{1},...,V_{K}$ are predictor variables and $\hat{\alpha}$ and $\hat{\beta_{1}},...,\hat{\beta_{K}}$ are parameters to be selected. In football matches, since, in some sense, a home win is `greater' than a draw which is `greater' than an away win, from equation~\ref{eq:ord}, the model can be parameterised as 

\begin{equation} \label{eq:ord1}
\log(\frac{\hat{p_{h}}}{\hat{p_{d}}+\hat{p_{a}}})=\hat{\alpha}+\sum_{i=1}^{K} \hat{\beta_{i}}V_{i},
\end{equation}
and
\begin{equation} \label{eq:ord2}
\log(\frac{\hat{p_{h}}+\hat{p_{d}}}{\hat{p_{a}}})=\hat{\alpha}+\sum_{i=1}^{K} \hat{\beta_{i}}V_{i}
\end{equation}
where $\hat{p_{h}}$, $\hat{p_{d}}$ and $\hat{p_{a}}$ are the estimated probabilities of a home win, a draw and an away win respectively.  These are easily found by solving with respect to equations~\ref{eq:ord1} and~\ref{eq:ord2}.  Throughout this paper, least squares parameter estimates are used to select the regression parameters $\hat{\alpha}$ and $\hat{\beta_{1}},...,\hat{\beta_{i}}$. Combinations of the following predictor variables are used: 
\begin{itemize}
\item The home team's odds-implied probability of winning.
\item Observed differences in the number of shots on target, shots off target and corners achieved by each team.
\item Differences in the predicted number of shots on target, shots off target, corners and goals for each team.
\end{itemize}
The home team's odds-implied probability is included as an extra predictor variable, in order to assess the importance of match statistics both individually and when used alongside the other information reflected in the odds. \par 

\subsection{Betting Strategies} \label{section:betting_strategy}
Following \cite{Wheatcroft2020}, in this paper, forecasts are constructed and used alongside two betting strategies: a simple level stakes value betting strategy and a strategy based on the Kelly Criterion. These are both described below. \par

Under the \emph{Level stakes} betting strategy, a unit bet is placed on the ith outcome of an event when $\hat{p}_{i}>r_{i}$, where $\hat{p}_{i}$ and $r_{i}$ are the predicted probability and the odds-implied probability respectively.  The idea is that a bet should be placed if the bet offers `value', that is if, in expectation, the bet would offer a positive return.  \par

The \emph{Kelly strategy}, is based on the Kelly Criterion (\cite{kelly1956new}) and has been used in, for example, \cite{Wheatcroft2020} and \cite{boshnakov2017bivariate}.  Under this approach, the amount staked on a bet is dependent on the difference between the forecast probability and the odds implied probability.  When the discrepancy between the forecast probability and the odds-implied probability is high, a greater amount of money is staked. Under the Kelly Criterion, the proportion of wealth placed on a bet for a particular outcome is given by
\begin{equation} \label{eq:kelly}
f_{i}=\max\left(\frac{r_{i}+{i}p_{i}-1}{r_{i}-1},0\right)
\end{equation}
where $p_{i}$ is the estimated probability of the outcome and $r_{i}$ represents the decimal odds on offer.  Under the Kelly Strategy used in this paper, the stake does not depend on the bank but is given by $s_{i}=kf_{i}$ where $k$ is a normalising constant set such that $\frac{1}{m} \sum_{i=1}^{m} kf_{i}=1$, where $m$ is the total number of bets placed.  The normalising constant is included purely to make the profit/loss from the Kelly Strategy directly comparable with that of the Level Stakes strategy. \par

Both the Level Stakes and Kelly betting strategies focus on the concept of `value' in which bets are only taken if the forecast implies a positive expected return.  It should be noted, however, that the two strategies are only guaranteed to find bets with value if the estimated probability and the true probability coincide.  In practice, due to model error in the forecasts, this can never be expected to be the case and the performance of the strategies must therefore be assessed empirically. \par

\subsection{Calculating GAP ratings} \label{section:gap_ratings}
In the GAP rating system, a team's ratings are updated each time it is involved in a match.  As such, its ratings after a match are dependent both on its previous ratings and its match performance. However, this leaves open the question of how to initialise the ratings of each team.  Whilst there are a number of approaches that could be taken, in the first season in which match statistics are available in a particular league, all GAP ratings are initialised at zero.  For subsequent seasons, a team's ratings are retained from one season to the next if they remain in the same league.  Teams relegated to a league are assigned the average ratings of those teams that were promoted in the previous season and teams that are promoted are assigned the average ratings of those teams that were relegated in the previous season. \par

For each match statistic, parameter selection is performed simultaneously over all leagues and takes place between seasons such that, at the beginning of each season, optimisation is performed over all previous seasons in which the relevant statistics are available.  Those parameters are then used for the entirety of the season (see section~\ref{section:parameter_selection} for details of how parameter selection is performed). The first season in which match statistics are available for any of the considered leagues (2000/2001) is used only for parameter selection purposes and therefore is not used for assessing the performance of the forecasts. \par

\section{Experiment}

\subsection{Experimental design}
The following experiment aims to assess the performance of observed and predicted match statistics in the forecasting of match outcomes.  This is done both in the context of traditional variable selection (using model selection techniques) and betting performance.  Following \cite{Wheatcroft2020}, the first six matches played by the home team in each season are not treated as eligible either for betting or for variable selection.  This is so that the model can have sufficiently `learned' about the quality of teams in that season. In addition, the final six matches played by the home team are also treated as ineligible since matches at this time of the season can be unpredictable due to differences in motivation (teams fighting for promotion/relegation etc). \par

In order to avoid overlap in the data, shots on target, shots off target, corners and goals are considered as potential predictor variables for the outcome of each match. The number of shots off target is calculated by subtracting the number of shots on target from the total number of shots.  Different combinations of observed and predicted match statistics are then assessed both with and without the odds-implied probability of the home team (calculated using the maximum odds over all bookmakers) included as an extra predictor variable.  \par

The first focus of the experiment is on variable selection and the aim is to find the combination of (i) observed and (ii) predicted match statistics that explain the match outcomes most effectively. This is done using Akaike's Information Criterion (AIC), which weighs up the fit of the model to the data with the number of parameters selected in-sample. As required for the calculation of information criteria, one set of regression parameters is selected over the entire data set. \par

Focus is then turned onto the question of betting performance.  Here, it is assumed that a gambler is able to `shop around' different bookmakers and take advantage of the highest odds offered on each outcome.  The maximum odds over all available bookmakers are thus assumed to be obtainable (note that the actual bookmakers included in the data set varies over time).  The forecasts are based only on information that would have been available at the time and are therefore produced truly `out-of-sample'.  New regression parameters are selected on each day in which at least one match is played and are calculated based on all eligible past matches.  The performance of the forecasts when used in conjunction with the Level Stakes and Kelly betting strategies defined in section~\ref{section:betting_strategy} is then assessed. Note that bets placed on draws are not considered due to the inherent difficulty of predicting them and therefore only bets on home or away wins are allowed. \par

\subsection{Evaluation of GAP ratings}
\begin{figure}[!htb]
    \centering
    \includegraphics[scale=0.7]{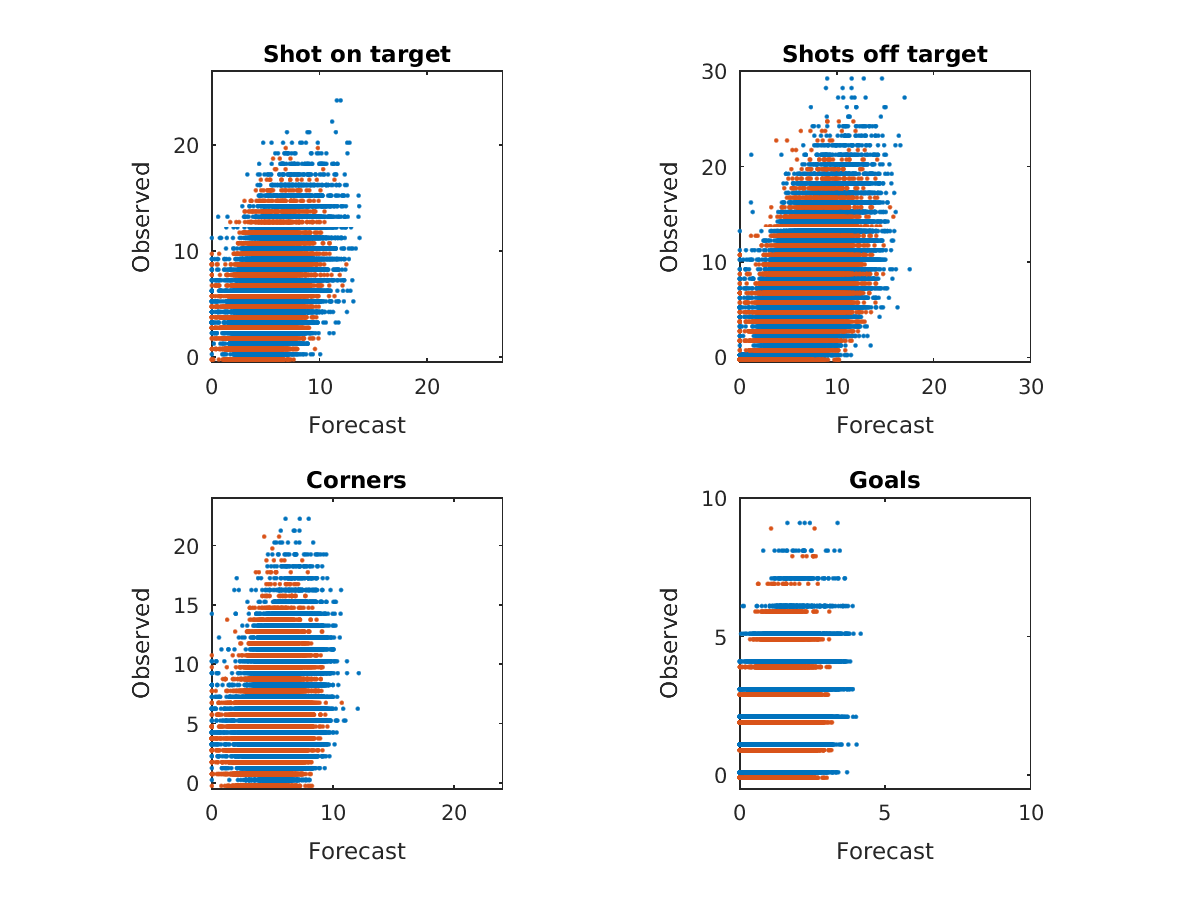}
    \caption{Scatterplots of predicted and observed match statistics.  The blue and red points correspond to home and away teams, respectively.}
    \label{figure:Scatterplots_shots_estimates}
\end{figure}

First, the performance of GAP ratings in terms of predicting match statistics is assessed. This is done by calculating the mean absolute error between the predicted and observed statistics.  For comparison, the mean absolute error between the observed statistic and the sample mean is calculated to highlight forecast performance when using the sample mean as a simple alternative forecast which requires no information about the relative strengths of the teams.  These are shown in table~\ref{table:deviations_stats} for each statistic, both for home and away teams. Here, there is a high degree of variation in the performance of the GAP ratings as predictors of the match statistics.  For the number of shots off target, the GAP ratings are particularly informative, strongly outperforming the sample mean.  For the other statistics, however, the improvement is much smaller.  Scatter plots of predicted statistics against their observed values are shown in figure~\ref{figure:Scatterplots_shots_estimates} for home (blue) and away (red) teams.  Variation in the performance of the GAP ratings is important in terms of the overall conclusions of this paper and is discussed later.  \par

\begin{table}[!htb] 
\fontsize{12}{12}\selectfont
\begin{center}
\begin{tabular}{|l|cc|} 
\hline
Measure & GAP rating & Mean forecast \\ 
\hline
Home Goals & $1.01$ & $1.02$ \\
Home Shots on target & $4.77$ & $5.22$ \\
Home Shots off target & $1.86$ & $3.77$ \\
Home Corners & $2.31$ & $2.34$ \\
\hline
Away Goals & $0.87$ & $0.85$ \\
Away Shots on target & $3.09$ & $3.24$ \\
Away Shots off target & $1.63$ & $3.09$ \\
Away Corners & $2.05$ & $2.08$ \\
\hline
\end{tabular}
\caption{Mean absolute error between observed and predicted match statistics along with that between observed statistics and their sample means.}
\label{table:deviations_stats}
\end{center}
\end{table}

\subsection{Variable selection: observed match statistics}
The results of variable selection when using observed match statistics are shown in table~\ref{table:obs_data_AIC}.  Here, the AIC for different combinations of statistics is shown both with and without the home odds-implied probability included as an additional predictor variable.  The AIC in each case is expressed relative to that of the ordinal regression model fitted without predictor variables (often known as the null model and denoted as `A0'), which simply outputs a constant set of probabilities for each match. The lower the AIC, the more support for that particular combination of variables.  \par

The results yield a number of conclusions.  The best AIC is achieved when the model includes all three observed match statistics both when the home odds-implied probability is included as an additional predictor variable and when it is not.  That the number of shots on target should have an impact on the match result should not come as a surprise, since all goals other than own goals and highly unusual events (such as the ball deflecting off the referee or, in one case in 2009, a beachball) result from a shot on target.  Interestingly, however, the inclusion of the number of corners and shots off target, which don't usually directly result in goals, improves the model even once shots on target are considered.  \par

It is also interesting to compare the effects of each observed match statistic as an individual predictor variable.  Unsurprisingly, the number of shots on target provides the most information, followed by corners and shots off target.  Interestingly, shots off target and corners do not provide much information when considered individually but add a great deal of information when combined with the number of shots on target and/or the home odds-implied probability.  It is a property of generalised linear models that some predictor variables are only informative in combination with other predictor variables and this appears to be the case here. \par

Finally, all three match statistics add information even when the odds-implied probability is included in the model. This is perhaps not surprising since match statistics give an indication of how the match \emph{actually} went. \par

In practice, of course, observed statistics are never available pre-match.  Despite this, the results shown here have important implications. Match statistics can be predicted and, if those predictions are informative enough, it stands to reason that skillful forecasts of the outcome of the match can be made. \par  

\begin{table}[!htb] 
\fontsize{8}{8}\selectfont
\begin{center}
\begin{tabular}{|l|ccc|rr|} 
\hline
Comb. variables & On target & Off target & Corners & AIC w/o odds & AIC w. odds \\ 
\hline
A1  & * & * & * & $-11646.1$ & $-14902.2$ \\ 
A3  & * &   & * & $-11404.1$ & $-14229.6$ \\ 
A2  & * & * &   & $-10358.1$ & $-13032.3$ \\ 
A4  & * &   &   & $-9366.6$  & $-11155.5$ \\ 
A5  &   & * & * & $-18.6$    & $-6897.5$  \\ 
A6  &   &   & * & $-19.5$    & $-6542.3$  \\ 
A7  &   & * &   & $-7.0$     & $-6452.4$  \\ 
A0  &   &   &   & $0$        & $-5619.1$   \\ 
\hline
\end{tabular}
\caption{AIC of each combination of observed match statistics with and without the home odds-implied probability included as a predictor variable. Variables that are included are denoted with a star and, in each case, AIC is given relative to the model fitted with only a constant term.}
\label{table:obs_data_AIC}
\end{center}
\end{table}

\subsection{Variable selection: predicted match statistics}
The results of variable selection with predicted match statistics are shown in table~\ref{table:pred_data_AIC}.  Unsurprisingly, the AIC is generally higher than for the observed case, implying that the information content is lower.  Despite this, predicted match statistics are able to provide information regarding match outcomes, even when the home odds-implied probability is included in the model. This means that, on average, the predicted match statistics provide information beyond that contained in the odds-implied probabilities. \par

It is of interest to note the relative importance of the different predicted match statistics.  Consistent with the finding of \cite{Wheatcroft2020}, the predicted number of goals is a relatively poor predictor of the outcome of the match.  It is also notable that whilst, in the observed case, the number of shots on target provides the most information about the outcome of the match, in the predicted case, the predicted shots off target is the most informative.  At first, this seems counterintuitive.  However, it should be noted that the information in the prediction is dependent both on the impact of the observed statistic on the match and the quality of the prediction of that statistic. The findings shown in table~\ref{table:deviations_stats} suggest that GAP rating predictions of shots off target are more accurate than those of the other match statistics and this is the likely explanation. \par

Finally, it is notable that, when considered as individual predictor variables, the \emph{predicted} number of shots off target and corners outperform the equivalent \emph{observed} statistics.  Again, this seems counterintuitive but can probably be explained by the fact that the predicted values consider the performances of the teams over multiple past matches, gaining some information about the relative strengths of the two teams. \par

\begin{table}[!htb]
\fontsize{8}{8}\selectfont
\begin{center}
\begin{tabular}{|l|cccc|rr|rr}
\hline
Comb. variables & Goals & On target & Off target & Corners & AIC w/o odds & AIC w. odds \\ 
\hline
B1  &   & * & * & * & $-4246.9$ & $-5750.8$ \\ 
B9  & * & * & * & * & $-4950.3$ & $-5749.3$ \\ 
B2  &   & * & * &   & $-4180.5$ & $-5739.5$ \\ 
B10 & * & * & * &   & $-4911.0$ & $-5738.1$ \\ 
B5  &   &   & * & * & $-3216.6$ & $-5733.7$ \\ 
B13 & * &   & * & * & $-4871.4$ & $-5732.3$ \\ 
B11 & * & * &   & * & $-4853.4$ & $-5729.1$ \\ 
B3  &   & * &   & * & $-4165.4$ & $-5728.3$ \\ 
B15 & * &   & * &   & $-4780.3$ & $-5709.4$ \\ 
B7  &   &   & * &   & $-2586.3$ & $-5709.1$ \\ 
B12 & * & * &   &   & $-4735.8$ & $-5698.1$ \\ 
B6  &   &   &   & * & $-2708.7$ & $-5697.6$ \\ 
B4  &   & * &   &   & $-4008.7$ & $-5696.9$ \\ 
B14 & * &   &   & * & $-4695.1$ & $-5695.8$ \\ 
B0  &   &   &   &   & $0$       & $-5619.1$ \\ 
B8  & * &   &   &   & $-4324.4$ & $-5617.8$ \\ 
\hline
\end{tabular}
\caption{AIC of each combination of predicted match statistics with and without the home odds-implied probability included as a predictor variable.  Included variables are denoted with a star and each AIC is given relative to the regression model with only a constant term.}
\label{table:pred_data_AIC}
\end{center}
\end{table}

\subsection{Betting performance}
Having determined that predicted match statistics can provide informative probabilistic forecasts, attention is now turned to the performance of the forecasts alongside the two betting strategies defined in section~\ref{section:betting_strategy}.  The mean percentage profit obtained from the Level Stakes betting strategy when used alongside forecasts derived from each combination of predicted match statistics is shown in table~\ref{table:profit_level}, along with 95 percent bootstrap resampling intervals. The resampling intervals are presented to demonstrate the robustness of the profit and, if the interval does not contain zero, the profit can be considered to be statistically significant. \par 

It is clear from the results that including combinations of predicted match statistics as predictor variables tends to yield a profit. In addition, for all combinations, including the home odds-implied probability as an additional predictor variable yields an increase in profit.  In some cases, when the home odds-implied probability is included, the profit is significant, i.e. the bootstrap resampling interval does not include zero. Whilst caution is advised in comparing the precise rankings of different combinations of variables, the best performing combinations tend to include the predicted number of shots off target.  The predicted number of goals, on the other hand, tends to have limited value.  When individual predicted statistics are considered, the ranking of the results is consistent with the variable selection results of table~\ref{table:pred_data_AIC} in that the best performing predicted variable is shots off target, followed by corners, shots on target and goals. \par

\begin{table}[!htb]
\fontsize{8}{8}\selectfont
\begin{center}
\begin{tabular}{|l|cccc|rr|rr}
\hline
Comb. variables & Goals & On target & Off target & Corners & Mean prof. w/o odds & Mean prof. w. odds \\ 
\hline
B5  &   &   & * & * & $+0.66 (-0.78,+2.51)$ & $+1.85 (+0.17,+3.39)$ \\ 
B9  & * & * & * & * & $+0.51 (-1.09,+2.06)$ & $+1.63 (+0.16,+3.23)$ \\ 
B2  &   & * & * &   & $+0.55 (-1.17,+2.15)$ & $+1.56 (+0.05,+3.17)$ \\ 
B1  &   & * & * & * & $+0.80 (-0.74,+2.70)$ & $+1.50 (-0.49,+2.97)$ \\ 
B13 & * &   & * & * & $+0.32 (-1.43,+1.70)$ & $+1.49 (-0.14,+3.31)$ \\ 
B10 & * & * & * &   & $+0.15 (-1.54,+1.79)$ & $+1.48 (+0.15,+3.48)$ \\ 
B11 & * & * &   & * & $+0.03 (-1.59,+1.60)$ & $+1.12 (-0.22,+3.02)$ \\ 
B7  &   &   & * &   & $-0.41 (-2.15,+1.41)$ & $+0.99 (-0.41,+2.43)$ \\ 
B3  &   & * &   & * & $+0.38 (-1.15,+2.26)$ & $+0.87 (-0.64,+2.30)$ \\ 
B6  &   &   &   & * & $-0.69 (-2.53,+1.06)$ & $+0.83 (-0.64,+2.46)$ \\ 
B14 & * &   &   & * & $-0.37 (-2.13,+0.93)$ & $+0.67 (-0.91,+2.34)$ \\ 
B15 & * &   & * &   & $-0.62 (-2.39,+1.14)$ & $+0.58 (-0.67,+2.16)$ \\ 
B12 & * & * &   &   & $-0.89 (-2.38,+0.58)$ & $+0.53 (-1.15,+2.22)$ \\ 
B4  &   & * &   &   & $-0.33 (-2.11,+1.43)$ & $+0.13 (-1.51,+1.80)$ \\ 
B0  &   &   &   &   & $-2.33 (-3.92,-0.39)$ & $-1.07 (-2.57,+1.03)$ \\ 
B8  & * &   &   &   & $-2.79 (-4.51,-1.41)$ & $-1.63 (-3.15,+0.06)$ \\ 
\hline
\end{tabular}
\caption{Mean percentage profit of Level Stakes strategy with each combination of predicted match statistics with and without odds-implied probabilities included as a predictor variable.  Included variables are denoted with a star}
\label{table:profit_level}
\end{center}
\end{table}

The mean profit obtained from using the forecasts alongside the Kelly strategy are shown in table~\ref{table:profit_kelly}. Here, notably, the mean profit is generally substantially higher than that achieved using the Level Stakes strategy.  Again, including the home odds-implied probability as an additional predictor variable yields improved results for all combinations of variables.  In fact, the profit is significant in all cases in which at least one predicted match statistic other than the number of goals is included alongside the home odds-implied probability. \par

\begin{table}[!htb] 
\fontsize{8}{8}\selectfont
\begin{center}
\begin{tabular}{|l|cccc|rr|rr}
\hline
Comb. variables & Goals & On target & Off target & Corners & Mean prof. w/o odds & Mean prof. w. odds \\ 
\hline
B1  &   & * & * & * & $+3.58 (+1.34,+5.74)$  & $+5.01 (+3.38,+6.76)$ \\ 
B9  & * & * & * & * & $+1.95 (+0.04,+3.86)$  & $+5.01 (+3.32,+6.70)$ \\ 
B10 & * & * &   & * & $+1.77 (-0.19,+3.82)$  & $+4.82 (+3.09,+6.63)$ \\ 
B2  &   & * &   & * & $+3.38 (+1.20,+5.46)$  & $+4.79 (+2.92,+6.71)$ \\ 
B5  &   &   & * & * & $+2.73 (+0.66,+4.69)$  & $+4.68 (+2.84,+6.53)$ \\ 
B13 & * &   & * & * & $+1.31 (-0.76,+3.36)$  & $+4.60 (+2.94,+6.44)$ \\ 
B11 & * & * & * &   & $+0.81 (-1.21,+2.90)$  & $+4.23 (+2.44,+5.93)$ \\ 
B7  &   &   &   & * & $+1.93 (-0.15,+3.97)$  & $+4.20 (+2.53,+5.93)$ \\ 
B3  &   & * & * &   & $+2.81 (+0.67,+4.93)$  & $+4.18 (+2.38,+6.03)$ \\ 
B15 & * &   &   & * & $+0.75 (-1.21,+2.69)$  & $+4.16 (+2.47,+5.87)$ \\ 
B12 & * & * &   &   & $-0.03 (-2.06,+1.95)$  & $+3.15 (+1.25,+5.01)$ \\ 
B6  &   &   & * &   & $+0.91 (-1.42,+3.24)$  & $+3.10 (+1.27,+4.81)$ \\ 
B14 & * &   & * &   & $-0.70 (-2.68,+1.28)$  & $+3.04 (+1.26,+4.88)$ \\ 
B4  &   & * &   &   & $+2.05 (+0.03,+4.09)$  & $+3.00 (+1.13,+5.00)$ \\ 
B8  & * &   &   &   & $-3.61 (-5.66,-1.61)$  & $-1.27 (-3.59,+1.16)$ \\ 
B16 &   &   &   &   & $-3.30 (-5.90,-0.78)$  & $-0.90 (-3.08,+1.17)$ \\ 
\hline
\end{tabular}
\caption{Mean percentage profit from the Kelly strategy using forecasts based on each combination of predicted match statistics with and without the home odds-implied probability included as a predictor variable.  Included variables are denoted with a star.}
\label{table:profit_kelly}
\end{center}
\end{table}

\begin{figure}[!htb]
    \centering
    \includegraphics[scale=0.7]{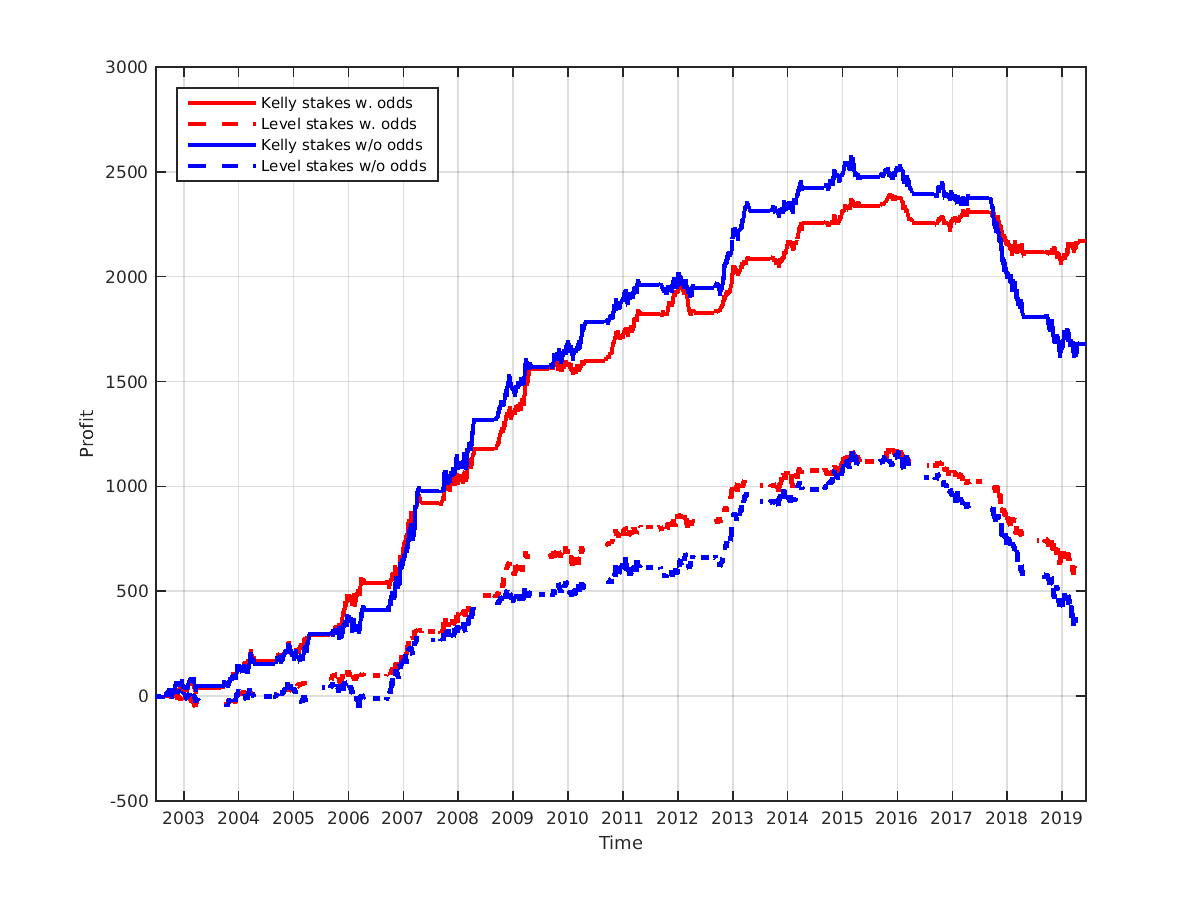}
    \caption{Cumulative profit from using the Kelly strategy (solid lines) and the level stakes betting strategies (dashed lines) when the home odd-implied probability is included as a predictor variable in the model (blue) and when it is excluded (red).}
    \label{figure:Profit_time_no_odds}
\end{figure}

For the remainder of this section, the betting performance of forecasts formed using predicted shots on target, shots off target and corners as predictor variables is considered, both with and without the home odds-implied probability as an additional predictor variable. The cumulative profit achieved with each of the two betting strategies is shown in figure~\ref{figure:Profit_time_no_odds}.  As already shown in tables~\ref{table:profit_level} and~\ref{table:profit_kelly}, a substantial profit is made in all four cases. The figure, however, shows how each strategy performs over time and an interesting feature is that there appears to be a downturn in profit in recent seasons.  Whilst this could conceivably be explained by random chance, it is perhaps more likely that something fundamental changed over that time. That predicted match statistics provide information additional to that contained in the odds suggests that, in general, the odds do not adequately account for the ability of teams to create shots and corners.  However, as more data have become available and quantitative analysis has become more sophisticated, it seems a reasonable claim that such information is now more likely to be reflected in the odds on offer and it may therefore be the case that the betting opportunities available in earlier seasons simply don't exist anymore. \par

\begin{figure}[!htb]
    \centering
    \includegraphics[scale=0.4]{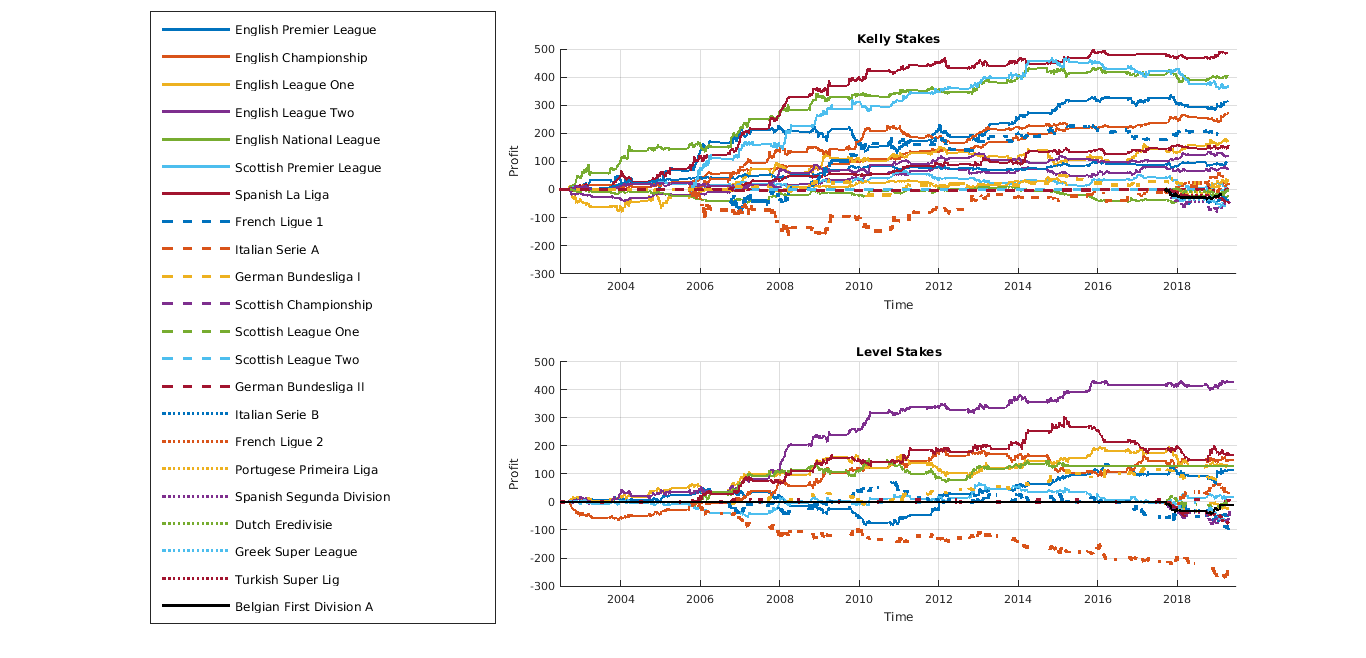}
    \caption{Cumulative profit as a function of time in each league for the case in which predicted shots on target, shots off target and corners along with the home odds-implied probability are included as predictor variables.}
    \label{figure:profit_over_time_by_league}
\end{figure}

It is worth considering how the profits from each betting strategy are distributed between the different leagues and whether losses in any particular subset of leagues can explain the observed downturn. Focusing on the case in which the home odds-implied probability is included as a predictor variable, in figure~\ref{figure:profit_over_time_by_league}, the cumulative profit made in each league is shown as a function of time.  Here, the decline in profit appears to be fairly consistent over all leagues considered and therefore, if the information reflected in the odds really has increased over time, this appears to be fairly universal over the different leagues. \par

\begin{figure}[!htb]
    \centering
    \includegraphics[scale=0.7]{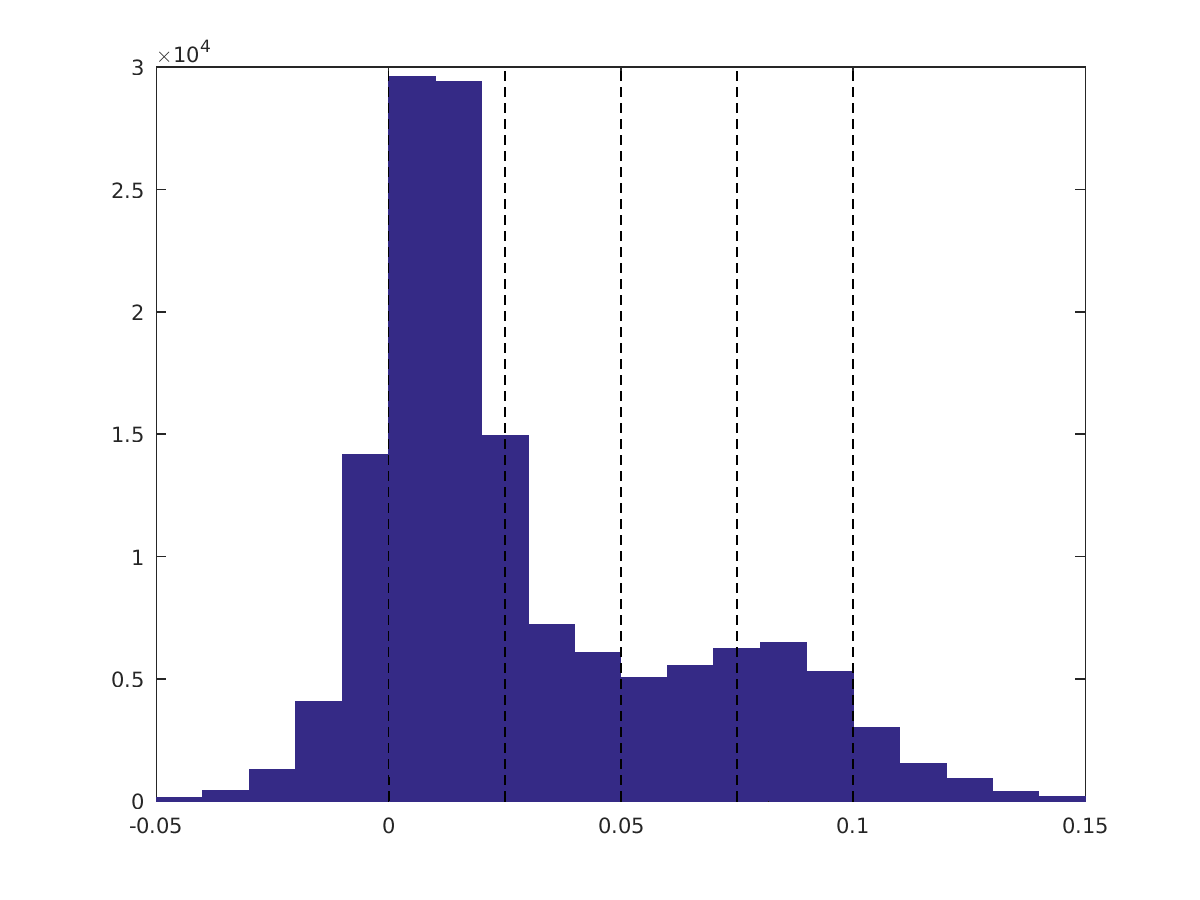}
    \caption{Histogram of overrounds for eligible matches when considering the maximum odds.}
    \label{figure:histogram_overrounds}
\end{figure}

Finally, it is important to assess the impact of the overround on the profitability of the betting strategies.  In this experiment, it is assumed that the gambler is able to find the best odds on offer on each possible outcome, over a range of bookmakers. Due to increased competition, there has been a trend towards reduced profit margins in recent years.  This can have a knock on effect on the overround of the best odds. A histogram of the overround of the best odds for all matches deemed eligible for betting is shown in figure~\ref{figure:histogram_overrounds}.  Whilst, in the majority of cases, the overround is positive, in around 18 percent of cases, it is negative. This gives rise to arbitrage opportunities, which means that a guaranteed profit can be made, without any need for a model. It is therefore important to distinguish cases in which profits are made due to the performance of the forecasts from those in which a profit could be guaranteed through arbitrage. \par

\begin{figure}[!htb]
    \centering
    \includegraphics[scale=0.7]{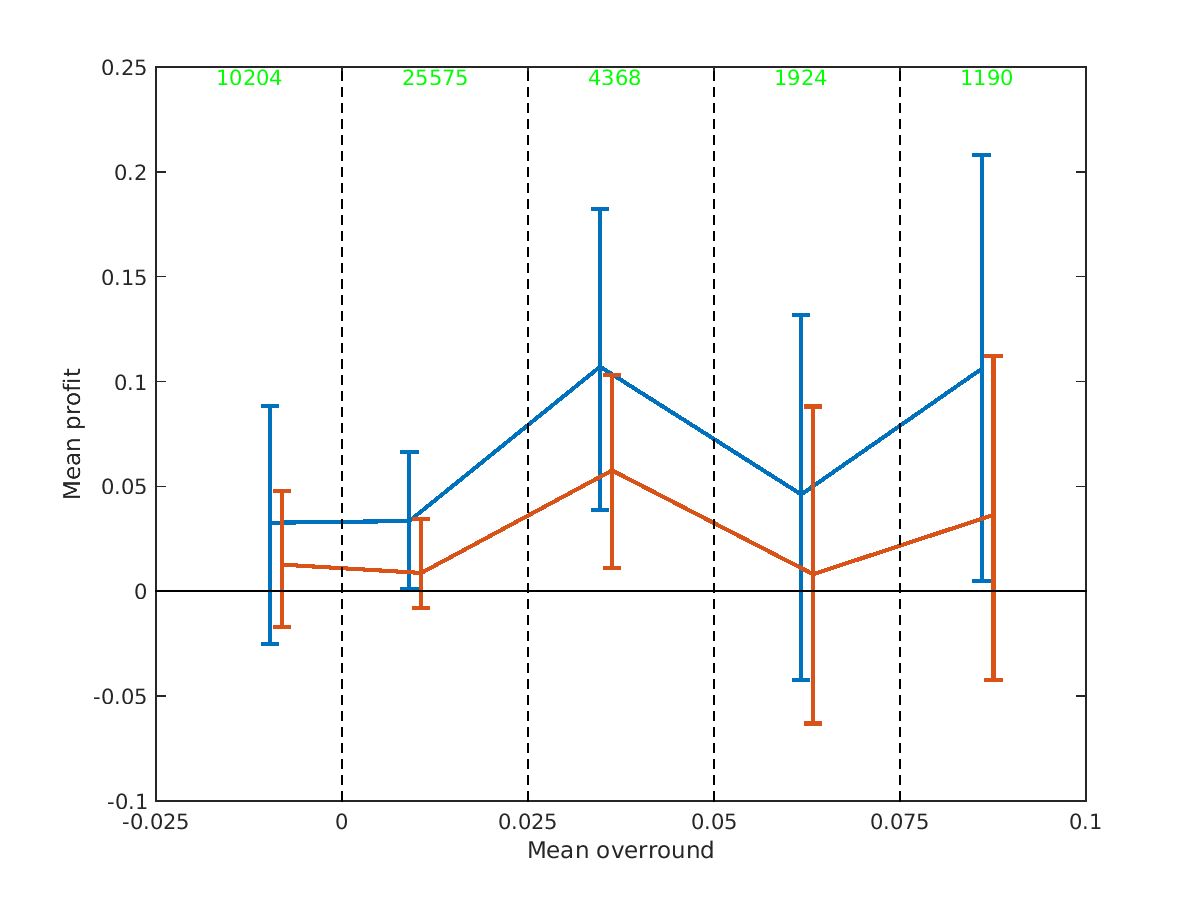}
    \caption{Mean overround against mean profit under the Kelly strategy (blue) and the Level Stakes strategy (red) for each considered interval. The error bars represent 95 percent bootstrap resampling intervals of the mean.}
    \label{figure:profit_function_overround}
\end{figure}

To assess the importance of the overround, five different intervals are defined and the mean profit from matches whose overround falls into each one is calculated under both betting strategies.  The first interval contains all matches with an overround less than zero, whilst, for matches with a positive overround, intervals with a width of 2.5 percent are defined.  The interval containing matches with the largest overrounds consider those in which the overround is greater than 7.5 percent. In figure~\ref{figure:profit_function_overround}, the mean overround for matches contained in each interval is plotted against the mean profit under each of the two betting strategies. The error bars correspond to 95 percent bootstrap resampling intervals of the mean profit.  In all five intervals, and under both betting strategies, the mean profit is positive. Under the Kelly strategy, three out of the five intervals yield a significant profit, whilst this is true in one case for the Level Stakes strategy.  Interestingly, the mean profit is not significantly different from zero when the overround is negative.  This, however, is consistent with the decline in profit in recent seasons that has tended to coincide with lower overrounds. Overall, the fact that significant profits can be made for matches in which the overround is positive suggest that, over the course of the dataset, the forecasts in combination with the two betting strategies would have been successful in identifying profitable betting opportunities. \par

\section{Discussion} \label{section:discussion}
In this paper, relationships between observed and predicted match statistics and the outcomes of football matches have been assessed. Unsurprisingly, the observed number of shots on target is a strong predictor of the match outcome whilst the observed numbers of shots off target and corners also provides some predictive value, once the number of shots on target and/or the match odds are taken into account.  With this in mind, the key claim of this paper is that \emph{predictions} of match statistics, if accurate enough, can be informative about the outcome of the match and, crucially, since the predictions are made in advance, this can aid betting decisions. \par

GAP ratings have been demonstrated to provide a convenient and straightforward approach to the prediction of match statistics. A number of interesting and, perhaps, surprising conclusions have been revealed.  Notably, in the prediction of match results, the most informative observed statistics do not coincide with the most informative predicted statistics. Whilst the number of shots on target were found to be the most informative observed statistic, the most informative predicted statistic was found to be the number of shots off target. As pointed out earlier in the paper, this can be explained by the fact that the information in the predicted statistics reflects both the importance of the statistic itself, in terms of the match outcome, and the accuracy of the prediction of that statistic. \par  

The observation above has interesting implications for the philosophy of sports prediction.  The importance of match statistics and, in particular, statistics such as expected goals that are derived from match events is becoming clear. The aim of expected goals can broadly be considered to be to estimate the expected number of goals a team `should' score, given the location and nature of the shots it has taken. A shot taken close to the goal and at a favourable angle has a high chance of being successful and therefore contributes more to a team's expected goals than a shot that is far away and from which it is difficult to score.  As such, expected goals ought to reflect the likelihood of each match outcome better than traditional statistics like the number of shots on target.  The results in this paper, however, suggest that it is not necessarily the case that the predicted number of expected goals by each team would outperform predictions of, or ratings based on, other statistics. Interesting future work would therefore be to predict the number of expected goals using GAP ratings and to assess the effect on the forecasting of match outcomes. \par

The results in this paper inspire a number of future avenues for research.  There is a wide and growing range of betting markets available for football matches and GAP ratings may be useful in informing such bets. This has already been shown by \cite{Wheatcroft2020} in the over/under 2.5 goal market but could also be applied to other markets such as Asian Handicap, the number of shots taken in a match, half time results and many more.  The philosophy demonstrated in this paper could also be applied to other sports. For example, in ice hockey, GAP ratings could be used to estimate the number of shots at goal, whilst, in American Football, they could be used to predict the number of yards gained by each team in the match. \par

Another interesting feature of the results presented in this paper is the decline in profit over the last few seasons. This was briefly discussed in the results section and it was suggested that betting odds now incorporate more information than at the beginning of the data set.  It would be interesting to investigate this further. \par

This paper demonstrates a new way of thinking about match statistics and their relationship with the outcomes of football matches and sporting events in general. It is hoped that this can help provide a better understanding of the role of match statistics in sports prediction and GAP ratings provide a straightforward and intuitive way in which to do this. \par

\newpage
\appendix

\section{GAP Ratings} \label{section:attack_defence_ratings}
GAP Ratings were introduced in a paper by \cite{Wheatcroft2020} aimed at predicting the probability that the total number of goals in a match will exceed $2.5$. They are defined as follows.  Consider a football league that contains $N$ different teams that each play one another several times (usually two) over the course of a season.  For a given match, let $S_{h}$ be a measure of the attacking performance of the home team and $S_{a}$ the equivalent for the away team.  The definition of `attacking performance' (referred to as the `model input' hereafter) is given by the user and, throughout this paper, is derived from match statistics.  In the paper in which the ratings were introduced, various model inputs are considered including goals, shots, shots on target and corners. \par

A separate attacking and defensive rating for each team for both home and away games defined as follows: 

\begin{itemize}
\item $H_{i}^{a}$ - The attacking rating of the ith team for home games
\item $H_{i}^{d}$ - The defensive rating of the ith team for home games
\item $A_{i}^{a}$ - The attacking rating of the ith team for away games
\item $A_{i}^{d}$ - The defensive rating of the ith team for away games
\end{itemize}
The attacking ratings can be considered to relate roughly to the attacking performance a team can be expected to achieve against an average team in the league.  The defensive ratings relate to the expected attacking performance of the average opposing team in the league.  Better teams will therefore have a high attacking rating and a low defensive rating.  After a match involving the ith team at home to the jth team, the GAP ratings for the ith team are updated in the following way
\begin{equation} 
\begin{split}
H_{i}^{a} & = \max(H_{i}^{a}+\lambda\phi_{1}(S_{h}-\frac{H_{i}^{a}+A_{j}^{d}}{2}),0), \\
A_{i}^{a} & = \max(A_{i}^{a}+\lambda(1-\phi_{1})(S_{h}-\frac{H_{i}^{a}+A_{j}^{d}}{2}),0), \\
H_{i}^{d} & = \max(H_{i}^{d}+\lambda\phi_{1}(S_{a}-\frac{A_{j}^{a}+H_{i}^{d}}{2}),0), \\
A_{i}^{d} & = \max(A_{i}^{d}+\lambda(1-\phi_{1})(S_{a}-\frac{A_{j}^{a}+H_{i}^{d}}{2}),0) \\
\end{split}
\end{equation}
The GAP ratings for the jth team are updated as follows:
\begin{equation} 
\begin{split}
A_{j}^{a} & = \max(A_{j}^{a}+\lambda\phi_{2}(S_{a}-\frac{A_{j}^{a}+H_{i}^{d}}{2}),0), \\
H_{j}^{a} & = \max(H_{j}^{a}+\lambda(1-\phi_{2})(S_{a}-\frac{A_{j}^{a}+H_{i}^{d}}{2}),0), \\
A_{j}^{d} & = \max(A_{j}^{d}+\lambda\phi_{2}(S_{h}-\frac{H_{i}^{a}+A_{j}^{d}}{2}),0), \\
H_{j}^{d} & = \max(H_{j}^{d}+\lambda(1-\phi_{2})(S_{h}-\frac{H_{i}^{a}+A_{j}^{d}}{2}),0), \\
\end{split}
\end{equation}
where $\lambda>0$, $0<\phi_{1}<1$ and $0<\phi_{2}<1$ are parameters to be selected. The parameter $\lambda$ represents the influence of a match on the ratings of each team whilst $\phi_{1}$ and $\phi_{2}$ decide the impact of a home match on a team's away ratings and of an away match on a team's home ratings respectively.  After any given match, a home team is said to have outperformed expectations in an attacking sense if its attacking performance is higher than the mean of its attacking rating and the opposition's defensive rating.  In this case, its home attacking rating is increased (or decreased, if its attacking performance is lower than expected).  If the parameter $\phi_{1}>0$, the home team's away ratings will be impacted whilst, the away team's home ratings will be impacted if $\phi_{2}>0$. \par

\newpage
\bibliographystyle{agsm}
\bibliography{bibliography}
\end{document}